\documentclass[conference]{IEEEtran}
\usepackage{cite}
\usepackage{amsmath,amssymb,amsfonts}
\usepackage{algorithmic}
\usepackage{graphicx}
\usepackage{textcomp}
\usepackage{tabularx}
\usepackage{array}
\usepackage{algorithm}
\usepackage{booktabs}
\usepackage{capt-of} 
\usepackage{microtype} 
\usepackage{balance} 
\usepackage{dblfloatfix} 

\def\BibTeX{{\rm B\kern-.05em{\sc i\kern-.025em b}\kern-.08em
    T\kern-.1667em\lower.7ex\hbox{E}\kern-.125emX}}
\begin{document}


\title{CARE-SD: Explainable Reasoning Architecture for Local SLM Self-Disclosure Analysis in Textual Data}

\author{\IEEEauthorblockN{Aisvarya Adeseye, Adeyemi Adeseye, Tahir Mohammad and Jouni Isoaho}
\IEEEauthorblockA{{Department of Computing, University of Turku, Turku, Finland}}
}

\maketitle

\begin{abstract}

Cybersecurity interviews uncover critical security experiences, behaviors, and privacy concerns often not fully captured by conventional technical assessments. Consequently, identifying genuine self-disclosures is challenging because statements are context-dependent, categories overlap, and automated interpretations may introduce unsupported claims. Locally deployed Small Language Models (SLMs) protect sensitive participant data by keeping processing within controlled environments, but their limited reasoning capabilities make simple prompting insufficient for context-aware and factually reliable analysis. This study developed CARE-SD (Context-Adaptive Reasoning and Explanation for Self-Disclosure), a novel four-module architecture integrating context-adaptive processing, hierarchical disclosure reasoning, explanation-consistency verification, and cross-batch reconciliation. We evaluated three local SLMs, including Llama 1B, Qwen 1.5B, and Llama 3B, using 82 independently annotated cybersecurity interviews across five organizational sectors. Comparisons included conventional classifiers, three prompting strategies, larger 8B--32B models, and architectural ablations. Two independent reviewers assessed disclosure labels, factual claims, and explanation correctness. CARE-SD improved the accuracy and evidence reliability of SLMs, increasing Llama 3B Macro-F1 from 0.80 to 0.88 while reducing unsupported claims by 60\%. Remarkably, the 3B model matched the 20B reference model in category Macro-F1, demonstrating the effectiveness of the architecture without requiring a larger model.

\end{abstract}


\section{Introduction}
\label{sec:introduction}

Organizational cybersecurity depends not only on technical safeguards but also on individuals' security experiences, practices, and perceptions of privacy and trust. Frameworks such as the NIST Cybersecurity Framework support risk management \cite{NIST2024}, yet technical assessments alone may overlook why employees adopt unsafe practices, experience incidents, or distrust monitoring. Qualitative interviews complement these assessments by capturing personal incidents, protective behaviors, and privacy concerns. Such insights are particularly relevant to gamified workplaces, where engagement mechanisms can introduce concerns about data collection and surveillance \cite{Adeseye2025,Adeseye2026TOPS}. Distinguishing what participants personally experienced, did, or feared can therefore inform security awareness, organizational policy, and privacy-conscious system design.

However, cybersecurity self-disclosure analysis is more demanding than identifying security-related keywords. Participants may implicitly describe phishing incidents, refer to earlier interview questions, combine several personal concerns, or discuss hypothetical organizational risks without disclosing anything personally. A response may contain overlapping disclosures of security experience (\textit{C1}), security/privacy behavior (\textit{C2}), and security/privacy concern (\textit{C3}). Failure to distinguish these categories can mischaracterise participants' risk exposure or inflate the apparent prevalence of behaviors and incidents. Earlier language-model research established fine-grained self-disclosure detection and taxonomies \cite{Dou2024}, but these findings do not directly establish accurate multi-label reasoning over context-dependent cybersecurity interview responses.

Large Language Models (LLMs) provide contextual interpretation, semantic classification, and human-readable explanations. Prior local-LLM studies involving cybersecurity interviews established the feasibility of qualitative privacy analysis \cite{Adeseye2025} and investigated methods for reducing hallucination \cite{Adeseye2026TOPS}. Nevertheless, fluent model-generated interpretations may attribute incidents to participants who discussed only hypothetical situations, infer unsupported motives, or present plausible explanations unsupported by transcript evidence. Such errors undermine qualitative cybersecurity conclusions, consistent with broader findings on generative hallucination \cite{Ji2023}. Consequently, disclosure analysis requires more than plausible category predictions; generated claims and explanations must be assessed against their sources.

Local deployment offers an important way to process sensitive transcripts within researcher-controlled computing environments rather than routinely transmitting them to external inference services. Furthermore, large LLMs require substantial computing resources. Small Language Models (SLMs) of approximately 1B--3B parameters offer a more accessible alternative, as demonstrated in \cite{Abdin2024} work on small capable language models. However, their restricted capacity can impair contextual reasoning, multi-label decisions, and explanation fidelity. Language models also use long contexts inconsistently \cite{Liu2024}; quantized-model studies have identified further reasons to introduce controlled verification \cite{Adeseye2026Quant}. Simply asking an SLM to detect disclosures, resolve references, assign overlapping categories, and justify each conclusion in one prompt leaves these subtasks entangled. Chain-of-thought prompting and iterative refinement help with reasoning \cite{Wei2022,Madaan2023}, but neither guarantees faithful evidence selection, and explanation quality must be independently assessed \cite{DeYoung2020}.

This study develops a modular and explainable architecture for local SLM-based cybersecurity self-disclosure analysis. The following research questions guided this work:

\textit{RQ1:} To what extent does the proposed architecture improve self-disclosure detection and category classification compared with conventional classifiers and standard SLM prompting approaches?

\textit{RQ2:} How effectively does the proposed architecture reduce unsupported interpretations and improve explanation accuracy while narrowing the performance gap with larger reference models?

The principal contribution of this study is CARE-SD, which integrates context-adaptive batch processing, hierarchical disclosure reasoning, explanation consistency verification, and cross-batch reconciliation. By separating SLM-based interpretation from deterministic context selection, validation, and aggregation, CARE-SD enables independently assessable predictions and supporting explanations. This study further establishes the classification and factual-reliability gains obtainable from this controlled architecture across locally deployable SLMs, offering a reproducible approach to evidence-grounded analysis under resource constraints.

\section{Related Work}
\label{sec:related_work}

\textbf{Detection and contextual classification:}
Self-disclosure detection has been studied in medical posts \cite{Valizadeh2023}, social media and annotated spans \cite{Dou2024,Yang2024,Lozoya2024,Luo2025}, and modality-dependent disclosures \cite{Grau2025,Henderson2026}. Dou \emph{et al.} \cite{Dou2024} developed a 19-category taxonomy, while Yang \emph{et al.} \cite{Yang2024} studied contextual extraction of rationales. These advances address disclosure identification, but their evaluation units and categories differ from question-dependent, overlapping \textit{C1--C3} disclosures in cybersecurity interviews. Particularly, an extracted span does not by itself establish the correctness of a generated interpretation.

\textbf{Privacy-sensitive text processing:}
Studies of personal disclosures in human--LLM conversations \cite{Mireshghallah2024}, context-aware privacy assistance \cite{Du2024,Krsek2025}, and sensitive-information detection \cite{Alfieri2024,Zeng2025} demonstrate important privacy applications. Locally deployable prompt reformulation \cite{Ngong2025}, sanitization \cite{AlfieriSanitize2025}, synthetic disclosure data \cite{Jangra2025}, and analyses of online communities \cite{WangLu2025,Haq2025} address related but distinct tasks. Detecting sensitive phrases, supporting a user's disclosure decision, or rewriting text does not establish whether category explanations generated by locally deployed SLMs faithfully represent the underlying interview evidence.

\textbf{Prior Interview Analysis Research, Prior Foundations, and Research Gap.}
Research on self-disclosure in conversational AI \cite{Papneja2024}, elicitation by interview agents \cite{Hama2025}, and semantic/thematic interpretation \cite{Chiang2025} provides relevant context for conversational settings. Also, prior studies by Adeseye \emph{et al.} \cite{Adeseye2025} examined LLM-assisted cybersecurity/privacy interviews, hallucination mitigation in qualitative privacy analysis \cite{Adeseye2026TOPS}, and multi-pass verification for quantized models \cite{Adeseye2026Quant}, laying the foundation for this study. CARE-SD does not claim these underlying techniques or the interview setting as new. Its distinct contribution is the \emph{disclosure-specific integration}: context-selected inputs, hierarchical multi-label decisions, quotation/explanation checks, and cross-window reconciliation, assessed jointly for category accuracy and claim-level reliability. 

\section{Proposed CARE-SD Architecture}
\label{sec:architecture}

\begin{figure*}[t]
\centering
\includegraphics[width=\textwidth]{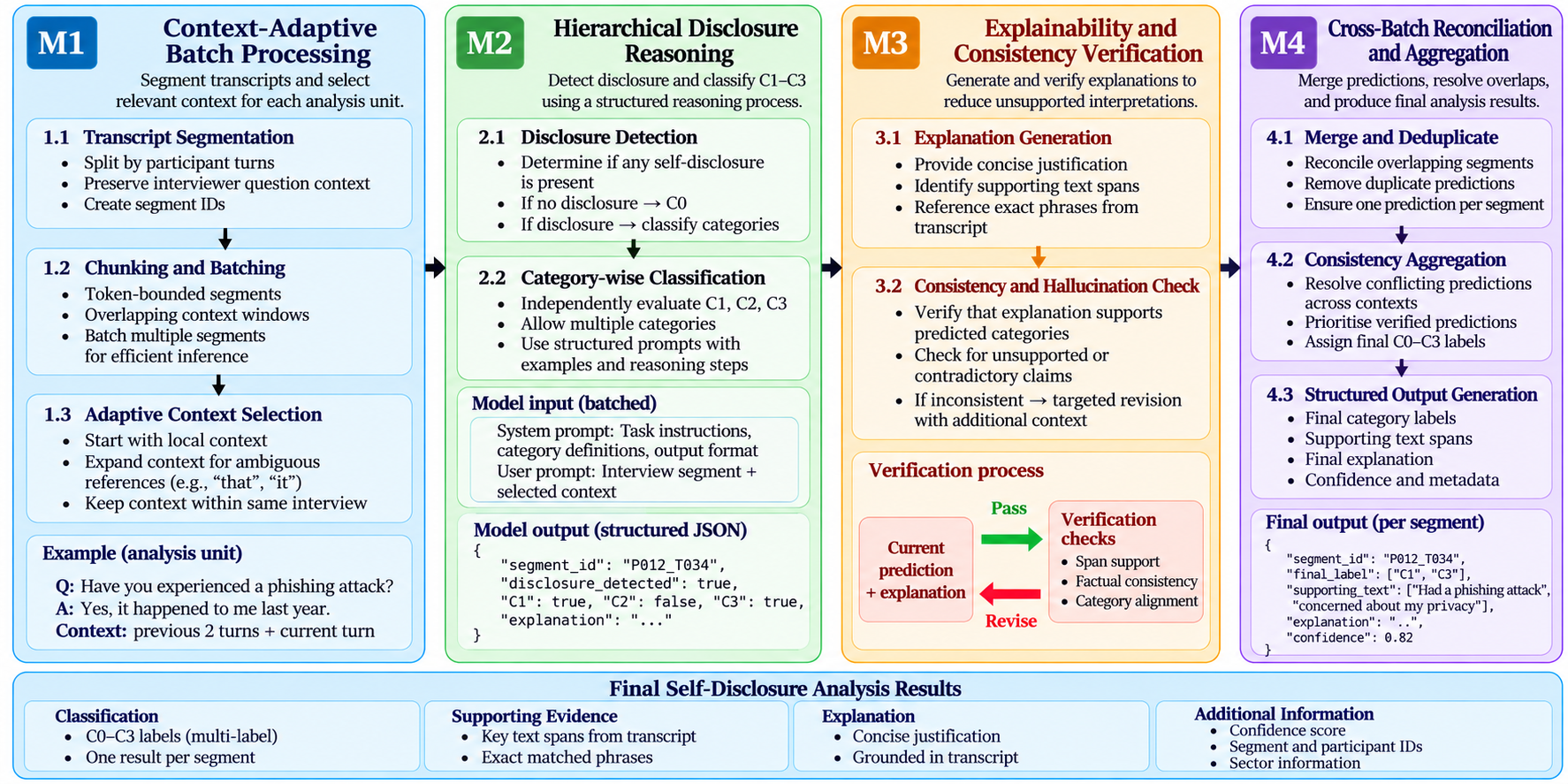}
\caption{The proposed four-module CARE-SD architecture for context-adaptive, explainable self-disclosure analysis using local SLMs. The four-module design separates semantic reasoning from deterministic context management, verification, and aggregation, reducing the complexity of individual SLM tasks while improving classification accuracy, evidence
reliability, and consistency across transcript segments.}
\label{fig:caresd_architecture}
\end{figure*}

This study developed CARE-SD, a four-module architecture designed to improve the accuracy and factual reliability of local SLMs in cybersecurity-related textual disclosure analysis. As illustrated in Fig.~\ref{fig:caresd_architecture}, the architecture comprises (M1) context-adaptive batch processing, (M2) hierarchical disclosure reasoning, (M3) explainability and consistency verification, and (M4) cross-batch reconciliation and aggregation. The design combined context-management principles \cite{Liu2024}, structured language-model reasoning \cite{Wei2022}, disclosure classification \cite{Dou2024}, and iterative verification \cite{Madaan2023}. Unlike unconstrained multi-pass prompting, CARE-SD separates SLM-based semantic interpretation from deterministic context selection, validation, and aggregation.

\subsection{Architectural Modules}

\textbf{M1--Context-Adaptive Batch Processing: }
Long transcripts were divided into sentence-aligned windows of $W=384$ tokens with $O=64$ tokens of overlap. Each window preserved the participant response, corresponding interviewer question, participant identifiers, and original text offsets. Up to $K=8$ independent windows were scheduled per inference batch without concatenating unrelated interviews. Because language models can overlook information within long contexts \cite{Liu2024}, the initial input was limited to the target participant response and its corresponding interviewer question. Context was expanded to the preceding $P=2$ question--answer pairs when predefined cues (``yes'', ``it'', ``that'', or ``as mentioned'') indicated contextual dependency or when M3 validation detected invalid output structure, inconsistent labels, or missing supporting quotations. A maximum input budget of $B=2048$ tokens was enforced to bound processing costs and limit irrelevant context while prioritizing the target participant response.

\textbf{M2--Hierarchical Disclosure Reasoning: }
Self-disclosure is decomposed into binary detection followed by independent category classification, reducing the complexity of a single multi-label decision. This design was motivated by explicit disclosure taxonomies \cite{Dou2024} and structured reasoning \cite{Wei2022}. For text window $i$, classification was defined as
\begin{equation}
\hat{y}_{ic}=\hat{d}_i\hat{z}_{ic},
\quad c\in\{1,2,3\},
\label{eq:classification}
\end{equation}
where $\hat{d}_i$ indicates disclosure presence and $\hat{z}_{ic}$ represents the predicted category: C1 (cybersecurity experiences), C2 (security/privacy behaviours), or C3 (personal security/privacy concerns). Multiple positive categories were permitted. C0 was assigned only when all category indicators were zero.

\textbf{M3--Explainability and Consistency Verification: }
The SLM generated concise classification explanations and verbatim supporting quotations. Inspired by rationale evaluation \cite{DeYoung2020} and iterative refinement \cite{Madaan2023}, a deterministic validator checked JSON conformity, permitted labels, category consistency, quotation existence, and explanation length. Each positive category required a supporting quotation from the participant response. When validation failed, M3 triggered at most one corrective re-analysis using the expanded conversational context provided by M1. Outputs remaining invalid after this retry were flagged rather than assigned the C0 non-disclosure label. These checks enforced structural and quotation consistency, while semantic correctness and unsupported interpretations were independently evaluated by human reviewers.

\textbf{M4--Cross-Batch Reconciliation and Aggregation: }
Overlapping windows were reconciled using participant identifiers and original text offsets. Duplicate quotations were removed, and valid category predictions were consolidated into one record per participant response:
\begin{equation}
\hat{Y}_{ic}=
\mathbb{I}\left(
\exists w\in\mathcal{W}_i:
v_w=1\land\hat{y}_{wc}=1
\right),
\label{eq:aggregation}
\end{equation}
where $\mathcal{W}_i$ represent the windows associated with response $i$, $v_w$ indicates successful validation, and $\mathbb{I}$ is the indicator function. This deterministic aggregation preserved valid disclosures across window boundaries. Responses with no valid predictions were flagged rather than automatically classified as non-disclosures.

\subsection{Algorithmic Implementation}

Algorithm~\ref{alg:caresd} presents the complete execution procedure. The controller explicitly defines segmentation, contextual triggers, inference order, validation conditions, retry limits, and aggregation. The SLM performs semantic detection and classification but does not control the processing workflow. All models used the same processing parameters and task definitions.

\begin{algorithm}[t]
\caption{CARE-SD Disclosure Analysis}
\label{alg:caresd}
\footnotesize
\begin{algorithmic}[1]

\REQUIRE Transcripts $T$, local SLM $f$;
$W=384$, $O=64$, $B=2048$, $P=2$, $K=8$
\ENSURE Consolidated disclosure records $Y$

\STATE $A\gets\mathrm{Segment}(T,W,O)$
\STATE Attach question, participant ID, offsets

\FOR{each batch $b\in\mathrm{Batch}(A,K)$}
  \FOR{each window $w\in b$}

    \STATE $x\gets\mathrm{Context}(w,B)$
    \STATE $y\gets\mathrm{Infer}(f,S_D,S_C,U(x))$

    \IF{$\mathrm{Cue}(w)$ OR $\neg\mathrm{Valid}(y,w)$}
      \STATE $x\gets\mathrm{Expand}(w,P,B)$
      \STATE $y\gets\mathrm{Infer}(f,S_D,S_C,U(x))$
    \ENDIF

    \IF{$\mathrm{Valid}(y,w)$}
      \STATE $\mathrm{Store}(w,y)$
    \ELSE
      \STATE $\mathrm{Store}(w,\mathrm{FLAG})$
    \ENDIF

  \ENDFOR
\ENDFOR

\STATE $Y\gets\mathrm{Reconcile}(A)$
\RETURN $Y$

\end{algorithmic}
\end{algorithm}

The inference function executed the detection system prompt $S_D$ first. When disclosure was detected, the classification system prompt $S_C$ independently assessed C1--C3; otherwise, C0 was assigned. The common user prompt $U$ supplied the segment identifier, interviewer question, selected preceding context, and participant response. Context expansion and retry decisions were executed by the controller rather than generated by the SLM.

\subsection{System and User Prompt Specification}

The architecture used two fixed system prompts and one common user template. The detection system prompt ($S_D$) instructs the model to determine whether the participant personally disclosed a cybersecurity experience, security/privacy behaviour, or personal security/privacy concern, excluding general observations and hypothetical statements. It returns a Boolean disclosure decision in JSON format.
The classification system prompt ($S_C$) defined C1--C3, permitted multiple categories, and required a verbatim quotation for each positive category and a concise explanation of at most 20 words. It explicitly prohibited invented experiences, motivations, or circumstances.
The user prompt followed the fixed structure
\begin{equation}
U_i=\{
\mathrm{ID}_i,\,
Q_i,\,
C_i,\,
R_i
\},
\label{eq:userprompt}
\end{equation}
where $\mathrm{ID}_i$ identifies the segment, $Q_i$ is the interviewer question, $C_i$ contains the controller-selected preceding context, and $R_i$ is the target participant response. Both system prompts required JSON-only outputs. All prompt instructions, token limits, and correction rules remained fixed across the evaluated SLMs.

\section{Experimental Setup}
\label{sec:experiments}

The experiments evaluated whether CARE-SD improved disclosure-category accuracy and reduced unsupported interpretation. The evaluation combined independently annotated interviews, conventional and prompting baselines, larger-model references, architectural ablations, and blinded output review.

\subsection{Dataset and Human Gold Standard}

The corpus comprised 82 completed researcher-led interviews on cybersecurity and privacy in gamified organizational environments. The participants represented NGOs ($n=35$), companies ($n=17$), academia ($n=11$), government ($n=10$), and healthcare ($n=9$). Interviews employed open-ended questions in private virtual meetings. Participant responses were segmented for analysis while retaining their corresponding questions as context. Two independent reviewers annotated the entire corpus using three potentially overlapping self-disclosure categories: C1, cybersecurity experiences and incidents; C2, security and privacy behaviours; and C3, personal security and privacy concerns. C0 indicated the absence of all three categories. Pre-adjudication Cohen's $\kappa$ values were $0.90$, $0.86$, and $0.88$ for C1--C3, respectively, yielding an unweighted mean of $\bar{\kappa}=0.88$. Disagreements were resolved through consensus to establish the human gold standard.
The dataset contained 1,482 annotated participant responses, partitioned into training (58 interviews; 1,028 responses), validation (12; 225), and held-out testing (12; 229). All responses from each participant were assigned to a single partition to prevent participant-level information leakage. Table~\ref{tab:dataset_stats} summarizes the distribution. C0 is exclusive, whereas C1--C3 may overlap within individual responses.

\begin{table}[t]
\caption{Dataset statistics and participant-disjoint partitions.}
\label{tab:dataset_stats}
\centering
\scriptsize
\setlength{\tabcolsep}{3pt}
\renewcommand{\arraystretch}{1.05}
\begin{tabular}{lrrrrrr}
\hline
\textbf{Split} & \textbf{Interviews} & \textbf{Annotated Responses} &
\textbf{C0} & \textbf{C1} & \textbf{C2} & \textbf{C3} \\
\hline
Training   & 58 & 1,028 & 240 & 360 & 450 & 420 \\
Validation & 12 & 225   & 53  & 80  & 95  & 88  \\
Test       & 12 & 229   & 56  & 81  & 98  & 85  \\
\hline
\textbf{Total} & \textbf{82} & \textbf{1,482} &
\textbf{349} & \textbf{521} & \textbf{643} &
\textbf{593} \\
\hline
\end{tabular}

\end{table}

Separately from gold-standard annotation, two reviewers, blinded to the model identity, independently assessed the generated outputs against the source transcripts. Each assessable factual claim was classified as supported, unsupported, or contradicted. The reviewers additionally evaluated whether the generated explanations correctly justified the assigned disclosure categories. Disagreements were resolved by consensus. Invalid outputs received one corrective retry. Predictions remaining invalid or flagged after reconciliation were retained in the held-out evaluation rather than excluded or automatically assigned C0. For detection and category F1-score, these predictions were counted as false negatives for gold-positive labels and false positives for gold-negative labels. Missing or invalid explanations were scored as incorrect whenever an explanation was required. The unsupported claim rate (UCR) and contradiction rate (CR) were computed as the respective proportions of unsupported and contradicted claims among all assessable generated factual claims. Outputs containing no assessable claims contributed nothing to these claim-level denominators and were recorded separately as flagged failures.

\subsection{Baselines and Reference Models}

Conventional baselines consisted of TF-IDF features with a linear support vector machine (SVM) \cite{Pedregosa2011} and a fine-tuned DeBERTa-v3-base classifier \cite{He2023}. They established lexical and contextual supervised reference performance, motivated by prior work on self-disclosure detection \cite{Dou2024}. Both used freely available, permissively licensed implementations.
The target SLMs were Llama-3.2-1B-Instruct, Qwen2.5-1.5B-Instruct, and Llama-3.2-3B-Instruct. The Llama models enabled within-family comparison across parameter scales, while Qwen provided cross-family evaluation. Each was evaluated under (i) direct zero-shot instruction, (ii) few-shot demonstrations from the training set \cite{Brown2020}, (iii) zero-shot step-by-step reasoning \cite{Kojima2022}, drawing on chain-of-thought prompting \cite{Wei2022}, and (iv) the proposed architecture. Larger, self-hosted references comprised Qwen3-8B, gpt-oss-20b, and Qwen3-32B. These supplied model-scale comparisons rather than ground-truth labels. 

\subsection{Inference and Quantization}

SLM inference used a local laptop with 64~GB of available memory. Larger models used a self-hosted AWS EC2 \texttt{g6e.2xlarge} instance with an NVIDIA L40S GPU (48~GB nominal VRAM) and 64~GiB system RAM. The specified configuration used 4-bit GGUF \texttt{Q4\_K\_M} weights for dense generative models and released native \texttt{MXFP4} weights for gpt-oss-20b. Within-model comparisons held the checkpoint, quantization, decoding, and task specification fixed. Hardware-dependent efficiency metrics were excluded because the environments differed.

\subsection{Evaluation Matrix and Protocol}

\textit{Evaluation procedure:} Table~\ref{tab:exp_matrix} summarises the 5 experiments, their measurement definitions, and computation procedures. Classification metrics were calculated against the adjudicated gold standard, whereas hallucination and explanation metrics used consensus judgments from two independent reviewers. Macro-F1 was calculated across the three disclosure categories (C1--C3), treating each as an independent binary label. C0 represented the absence of all disclosure categories. E1 and E2 established the primary accuracy and hallucination outcomes; E3 assessed individual component contributions; E4 examined model-scale differences; and E5 evaluated explanation correctness. 
Participant-disjoint splitting was used for point estimates. Confidence intervals are omitted because response-level paired predictions, participant identifiers, and claim adjudications required to independently rerun the reported bootstrap are not included in the available result records. Accordingly, differences are presented descriptively rather than as statistical-significance or equivalence findings.

\begin{table*}[!t]
\caption{Experimental objectives, metrics, and evaluation procedures.}
\label{tab:exp_matrix}
\centering

\footnotesize
\setlength{\tabcolsep}{4.5pt}
\renewcommand{\arraystretch}{1.30}
\setlength{\extrarowheight}{1pt}

\begin{tabularx}{\textwidth}{@{}
>{\raggedright\arraybackslash}p{.045\textwidth}
>{\raggedright\arraybackslash}p{.195\textwidth}
>{\raggedright\arraybackslash}p{.210\textwidth}
>{\raggedright\arraybackslash}X
@{}}

\toprule

\textbf{Exp.} &
\textbf{Objective} &
\textbf{Metrics} &
\textbf{Measurement and comparison}\\

\midrule

E1 &
\textbf{Architecture effectiveness:}
detect and classify disclosures. &
$F1=\frac{2TP}{2TP+FP+FN}$;
\par\vspace{3pt}
$MF1=\frac{1}{3}\sum_{c=1}^{3}F1_c$. &
Against adjudicated labels: binary disclosure
(C1--C3 versus C0) and mean binary F1 across
C1--C3; compare prompting strategies and
conventional classifiers.\\

\addlinespace[4pt]

E2 &
\textbf{Hallucination reduction:}
limit unsupported and contradictory claims. &
$UCR=U/N$;
\par\vspace{3pt}
$CR=C/N$;
\par\vspace{3pt}
$N=S+U+C$. &
Two blinded human reviewers adjudicated
generated claims as supported ($S$),
unsupported ($U$), or contradicted ($C$)
against source transcripts.\\

\addlinespace[4pt]

E3 &
\textbf{Component ablation:}
isolate module contributions. &
$MF1$; $UCR$
(defined above). &
Remove each module individually ($-j$);
compare with full CARE-SD, keeping the
SLM and test set fixed.\\

\addlinespace[4pt]

E4 &
\textbf{Model-scale comparison:}
assess smaller versus larger models. &
$MF1$; $UCR$
(defined above). &
Evaluate CARE-SD-enhanced 1B--3B SLMs
and 8B--32B references on identical
held-out samples.\\

\addlinespace[4pt]

E5 &
\textbf{Explanation correctness:}
assess the validity of explanations. &
$ECR=E_{\rm ok}/E_{\rm all}$. &
Two independent reviewers adjudicate
correct explanations ($E_{\rm ok}$)
among evaluated explanations ($E_{\rm all}$).
Sector-wise Macro-F1 was proposed but
not available for reporting.\\

\bottomrule

\end{tabularx}

\vspace{5pt}

\begin{minipage}{0.98\textwidth}
\footnotesize
\raggedright
\emph{Definitions:}
$TP$, $FP$, $FN$: true positives, false positives,
false negatives;
$F1_c$: binary F1 for category C1, C2, or C3;
$N$: total evaluable factual claims.
All claim/explanation outcomes use consensus
human-review judgments.
\end{minipage}

\end{table*}

\subsection{Experimental Comparisons}

\textit{E1} tested architectural accuracy gains against direct, few-shot, and step-by-step prompting in the same SLM. \textit{E2} measured hallucination separately because a correct category could still have an unsupported explanation. \textit{E3} removed one architectural component at a time to attribute changes in Macro-F1 and UCR to its contribution. \textit{E4} compared architecture-enhanced SLMs with larger references using the same labels and test set. \textit{E5} evaluated human-judged explanation correctness \cite{DeYoung2020}. 

\section{Results and Discussion}
\label{sec:results}

\begin{figure}
    \centering
    \includegraphics[width=0.8\linewidth]{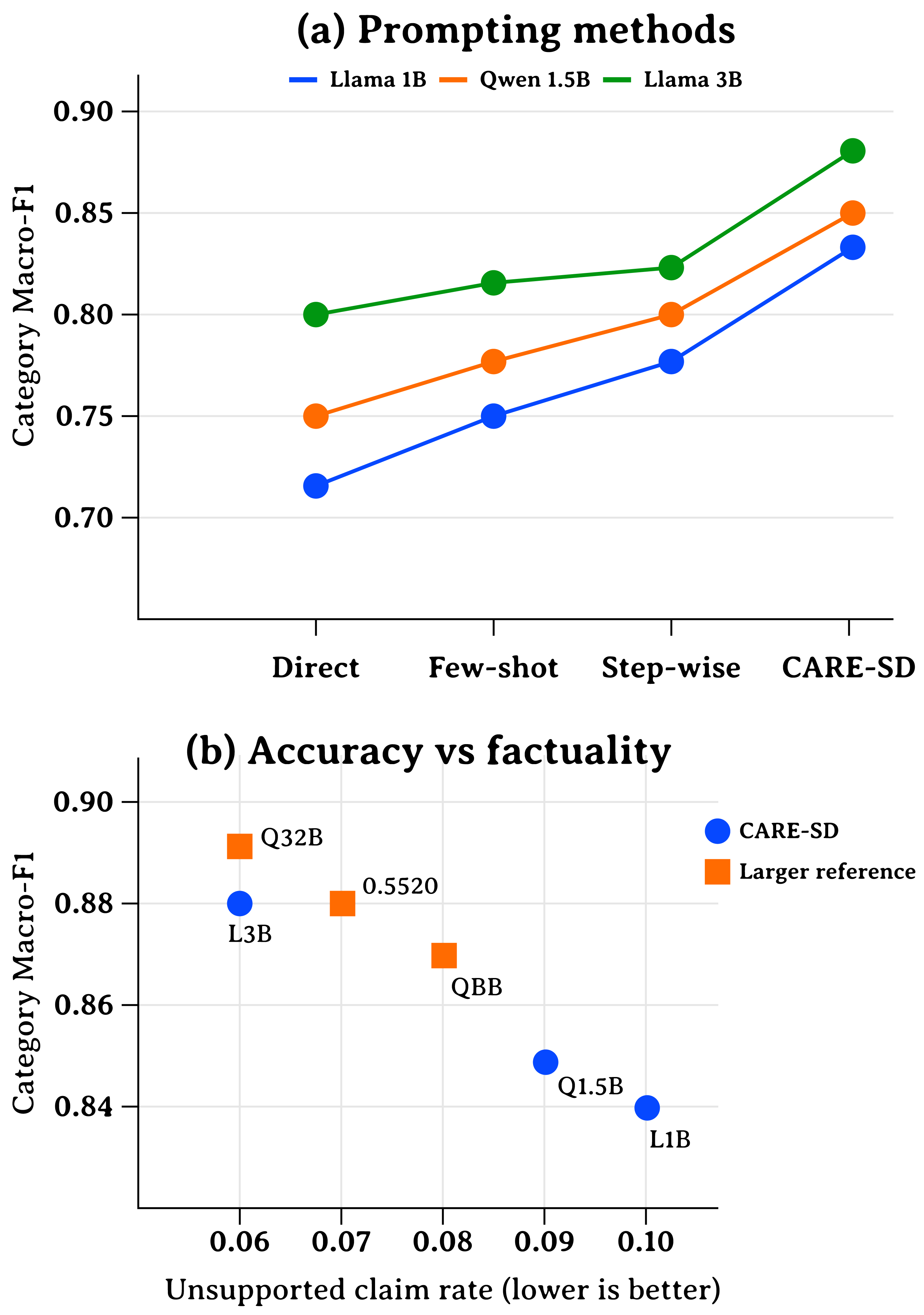}
    \caption{CARE-SD performance:
(a) category Macro-F1 across prompting approaches;
(b) Macro-F1 versus unsupported claim rate for
three SLMs and three larger reference models.
Ablation appears in
Table~\ref{tab:caresd_ablation}.}
    \label{fig:caresd_results}
\end{figure}

Table~\ref{tab:caresd_main_results} reports disclosure detection and category classification against the adjudicated gold standard, together with human-assessed factual reliability and explanation correctness. Detection F1 treated C1--C3 as disclosure and C0 as no disclosure; Macro-F1 was averaged over the three potentially overlapping categories. Unsupported Claim Rate (UCR) and Contradiction Rate (CR) were calculated over evaluable factual claims (lower is better), whereas Explanation Correctness Rate (ECR) measured human-judged correct explanations (higher is better). Figure~\ref{fig:caresd_results} summarizes prompting and model-scale comparisons; component ablations appear in Table~\ref{tab:caresd_ablation}.

\begin{table}[!t]
\centering
\caption{Disclosure classification and factual reliability.
$\uparrow$: higher is better; $\downarrow$: lower is better.}
\label{tab:caresd_main_results}

{\fontsize{6.5}{7.5}\selectfont
\setlength{\tabcolsep}{1.8pt}
\renewcommand{\arraystretch}{1.02}

\begin{tabularx}{\columnwidth}{@{}Xccccc@{}}
\hline
\textbf{Model / method} &
\textbf{Det. F1} &
\textbf{MF1} &
\textbf{UCR} &
\textbf{CR} &
\textbf{ECR}\\
\hline

\multicolumn{6}{@{}l}{\textit{Conventional classifiers}}\\
TF-IDF + SVM & .78 & .71 & -- & -- & --\\
DeBERTa-v3-base & .84 & .79 & -- & -- & --\\
\hline

\multicolumn{6}{@{}l}{\textit{Local small language models}}\\
Llama 1B / direct & .79 & .72 & .21 & .09 & .75\\
Llama 1B / few-shot & .81 & .75 & .19 & .08 & .78\\
Llama 1B / step-by-step & .83 & .78 & .16 & .07 & .81\\
\textbf{Llama 1B / CARE-SD} &
\textbf{.87} & \textbf{.84} & \textbf{.10} &
\textbf{.04} & \textbf{.88}\\

Qwen 1.5B / direct & .81 & .75 & .19 & .08 & .78\\
Qwen 1.5B / few-shot & .83 & .78 & .17 & .07 & .80\\
Qwen 1.5B / step-by-step & .84 & .80 & .15 & .06 & .83\\
\textbf{Qwen 1.5B / CARE-SD} &
\textbf{.88} & \textbf{.85} & \textbf{.09} &
\textbf{.03} & \textbf{.90}\\

Llama 3B / direct & .84 & .80 & .15 & .06 & .84\\
Llama 3B / few-shot & .85 & .82 & .14 & .05 & .86\\
Llama 3B / step-by-step & .86 & .83 & .12 & .05 & .88\\
\textbf{Llama 3B / CARE-SD} &
\textbf{.90} & \textbf{.88} & \textbf{.06} &
\textbf{.02} & \textbf{.93}\\
\hline

\multicolumn{6}{@{}l}{\textit{Larger reference models (CARE-SD)}}\\
Qwen3 8B & .90 & .87 & .08 & .03 & .92\\
gpt-oss 20B & .91 & .88 & .07 & .02 & .93\\
Qwen3 32B & .92 & .89 & .06 & .02 & .94\\
\hline

\end{tabularx}
}

\vspace{2pt}

\parbox{\columnwidth}{
\fontsize{6.5}{7.5}\selectfont
\raggedright
\emph{Note:} MF1 = category Macro-F1;
UCR = unsupported claim rate;
CR = contradicted claim rate;
ECR = explanation correctness rate.
Dashes mean not evaluated.
Claim counts $S/U/C$
(supported/unsupported/contradicted),
flagged-output count $F$, and explanation counts
$E_{\rm ok}/E_{\rm all}$ were not supplied
with the aggregate result records;
percentages cannot be independently recalculated
without these denominators.
\par
}

\end{table}

\subsection{E1: Architecture Effectiveness}

CARE-SD improved category Macro-F1 over direct prompting from 0.72 to 0.84 for Llama 1B (+0.12), 0.75 to 0.85 for Qwen 1.5B (+0.10), and 0.80 to 0.88 for Llama 3B (+0.08), as shown in Fig.~\ref{fig:caresd_results}(a). Detection F1 also increased to 0.87, 0.88, and 0.90, respectively (Table~\ref{tab:caresd_main_results}). The proposed configuration exceeded the strongest conventional baseline, DeBERTa-v3-base (Macro-F1 = 0.79), for all three SLMs. These comparisons isolated the gains within each underlying SLM, rather than attributing them to model scale alone.

\subsection{E2: Hallucination Reduction}

Unsupported claims decreased under CARE-SD for all three SLMs: UCR fell from 0.21 to 0.10 (Llama 1B), 0.19 to 0.09 (Qwen 1.5B), and 0.15 to 0.06 (Llama 3B). The corresponding CR values declined from 0.09 to 0.04, 0.08 to 0.03, and 0.06 to 0.02. For Llama 3B, this represented a 60.0\% relative UCR reduction (absolute difference: -0.09). Thus, gains in category classification were accompanied by fewer unsupported or contradicted statements in generated interpretations; the two outcomes were evaluated separately because correct labels could still have incorrect explanations.

\subsection{E3: Architectural Ablation}

Single-module ablations were evaluated using Llama 3B (Table~\ref{tab:caresd_ablation}). Full CARE-SD achieved Macro-F1 = 0.88 and UCR = 0.06. Removing M1, M2, M3, and M4 reduced Macro-F1 by 4, 5, 3, and 2 percentage points, respectively. The corresponding UCR increases were 5, 6, 9, and 3 percentage points. M2 contributed most strongly to classification accuracy, whereas removing M3 increased UCR from 0.06 to 0.15, representing the largest deterioration in factual reliability (+9 percentage points; 150\% relative increase). This finding highlights the contribution of quotation verification, consistency checking, and controlled re-analysis to limiting unsupported interpretations. However, the ablation did not isolate the individual effects of these verification operations or determine interactions between modules.

\begin{table*}[!t]
\caption{Component-wise ablation of CARE-SD with
Llama-3.2-3B-Instruct (same evaluation data;
MF1 higher and UCR lower are better).}
\label{tab:caresd_ablation}
\centering

\footnotesize
\setlength{\tabcolsep}{4pt}
\renewcommand{\arraystretch}{1.30}
\setlength{\extrarowheight}{1pt}

\begin{tabularx}{\textwidth}{@{}
>{\raggedright\arraybackslash}p{0.15\textwidth}
>{\raggedright\arraybackslash}X
ccccc
@{}}

\toprule

\textbf{Configuration} &
\textbf{Disabled architectural functionality} &
\shortstack{\textbf{Macro-F1}\\$\uparrow$} &
\shortstack{\textbf{F1 loss}\\\textbf{(pp)}$\downarrow$} &
\shortstack{\textbf{UCR}\\$\downarrow$} &
\shortstack{\textbf{UCR rise}\\\textbf{(pp)}$\downarrow$} &
\shortstack{\textbf{Relative}\\\textbf{UCR rise}} \\

\midrule

\textbf{Full CARE-SD} &
All four modules enabled: adaptive context,
hierarchical reasoning, explanation verification,
and cross-batch reconciliation &
\textbf{0.88} &
-- &
\textbf{0.06} &
-- &
-- \\

\midrule

Without M1 &
Context-adaptive input selection and
targeted conversational context expansion &
0.84 &
4 &
0.11 &
5 &
83.3\% \\

\addlinespace[5pt]

Without M2 &
Hierarchical disclosure detection followed
by independent multi-label classification
of C1--C3 &
0.83 &
\textbf{5} &
0.12 &
6 &
100.0\% \\

\addlinespace[5pt]

Without M3 &
Explanation-consistency verification,
supporting-quotation checks, and
targeted validation of model outputs &
0.85 &
3 &
0.15 &
\textbf{9} &
\textbf{150.0\%} \\

\addlinespace[5pt]

Without M4 &
Cross-window reconciliation,
overlap handling, and aggregation
of disclosure predictions &
0.86 &
2 &
0.09 &
3 &
50.0\% \\

\bottomrule

\end{tabularx}

\vspace{6pt}

\begin{minipage}{0.98\textwidth}
\footnotesize
\raggedright

\emph{Note:}
M1 = context-adaptive batch processing;
M2 = hierarchical disclosure reasoning;
M3 = explainability and consistency verification;
M4 = cross-batch reconciliation.

F1 loss =
$100(0.88-F1_{\mathrm{ablated}})$
percentage points.

UCR rise =
$100(UCR_{\mathrm{ablated}}-0.06)$
percentage points.

Relative UCR rise =
$100(UCR_{\mathrm{ablated}}-0.06)/0.06$
percent.

Each ablation removes one architectural module
while retaining the remaining components.

\end{minipage}

\end{table*}

\subsection{E4: Larger-Model Comparison}

Llama 3B + CARE-SD achieved Macro-F1 = 0.88, matching the reported two-decimal score of gpt-oss 20B, exceeding Qwen3 8B (0.87), and trailing Qwen3 32B (0.89) by one percentage point (Table~\ref{tab:caresd_main_results}). This numerical agreement does not establish statistical equivalence. Its UCR of 0.06 equaled Qwen3 32B and was lower than Qwen3 8B (0.08) and gpt-oss 20B (0.07), as shown in Fig.~\ref{fig:caresd_results}(b). Llama 1B and Qwen 1.5B also achieved Macro-F1 values of 0.84 and 0.85, respectively. Computational efficiency was not directly compared because inference hardware differed.

\subsection{E5: Explanation Correctness}

Human-assessed explanation correctness (ECR) increased from 0.75 to 0.88 for Llama 1B, 0.78 to 0.90 for Qwen 1.5B, and 0.84 to 0.93 for Llama 3B under CARE-SD compared with direct prompting. Llama 3B matched the gpt-oss 20B reference (0.93) and approached Qwen3 32B (0.94). These improvements complemented the reductions in unsupported claims, indicating that the complete architecture improved both explanation correctness and factual reliability. However, the deterministic verification in M3 established structural and quotation consistency rather than semantic correctness, which was assessed independently by human reviewers.

\subsection{Discussion and Limitations}

Across the reported comparisons, CARE-SD improved category classification and reduced unsupported interpretations without SLM fine-tuning. The ablations associated hierarchical reasoning most strongly with accuracy, and verification most strongly with factual reliability.
Nevertheless, the findings were specific to 82 cybersecurity-oriented interviews. The held-out test comprises only 12 interviews, and sector sizes were unequal. No independently reproducible confidence intervals or sector-specific scores were available from the supplied aggregates. The absence of claim-level numerators and denominators, flagged-output counts, and explanation-count denominators limits independent audit of UCR, CR, and ECR. Numerical equality in Macro-F1 between Llama 3B + CARE-SD and gpt-oss 20B does not establish statistical equivalence.  The joint accuracy--factuality comparison was not treated as a Pareto frontier because some reference configurations dominated others on both plotted metrics.

\section{Conclusion}
\label{sec:conclusion}

This study developed and evaluated CARE-SD, a four-module reasoning architecture for accurate and explainable self-disclosure analysis using locally deployed small language models. The architecture integrated  M1: Context-Adaptive Batch Processing, M2: Hierarchical Disclosure Reasoning, M3: Explainability and Consistency Verification, and M4: Cross-Batch Reconciliation to address contextual ambiguity and unsupported interpretations without modifying model parameters. Evaluation across 82 cybersecurity interviews demonstrated improvements in disclosure classification, factual reliability, and explanation correctness. With CARE-SD, Llama 3B improved category Macro-F1 from 0.80 to 0.88 and reduced the unsupported claim rate from 0.15 to 0.06, a 60\% reduction. It also achieved detection F1 of 0.90 and explanation correctness of 0.93, numerically matching the 20B reference model's Macro-F1. 
Ablation results identified M2 (Hierarchical Disclosure Reasoning) as the largest contributor to classification accuracy and M3 (Explainability and Consistency Verification) as the most influential module for reducing unsupported claims.

 These findings demonstrate that CARE-SD empowers resource-constrained SLMs to achieve classification performance comparable to a substantially larger model while improving factual reliability, strengthening their practical viability for privacy-sensitive self-disclosure analysis.
Future work will evaluate CARE-SD on independent, larger, and more diverse disclosure datasets to establish cross-domain generalizability and robustness.

\balance

\end{document}